\documentclass[aps,prl,showpacs,twocolumn,superscriptaddress]{revtex4} 
\newcommand{\bfi}[1]{\mbox{\boldmath $#1$}}
\newcommand{\Lower}[1]{\smash{\lower 1.5ex \hbox{#1}}}

\newcommand{\LNSN}{$\Lambda N-\Sigma N$}

\newcommand{\BL}{$B_\Lambda$}
\newcommand{\PS}{$P_\Sigma$}
\newcommand{\HII}{$^2$H}
\newcommand{\HIII}{$^3$H}
\newcommand{\HeIII}{$^3$He}
\newcommand{\HeIV}{$^4$He}
\newcommand{\HIIIL}{$_\Lambda^3$H}
\newcommand{\HIVL}{$_\Lambda^4$H}

\newcommand{\HeIVL}{$_\Lambda^4$He}

\newcommand{\HIVLs}{$_\Lambda^4$H$^\ast$}
\newcommand{\HeIVLs}{$_\Lambda^4$He$^\ast$}
\newcommand{\HeVL}{$_\Lambda^5$He}

\newcommand{\HeVILL}{$_{\Lambda\Lambda}^{\ \ 6}$He}

\usepackage{graphicx}

\begin{document} 				

\title{ 
Ab Initio Approach to $s$-Shell Hypernuclei \HIIIL, \HIVL, \HeIVL\ and \HeVL\\ 
with a \LNSN\ Interaction  \\ 
}
\author{H.~{Nemura}} 		
\author{Y.~{Akaishi}} 		
\affiliation{ 					
Institute of Particle and Nuclear Studies, KEK, Tsukuba 305-0801, Japan
} 						
\author{Y.~{Suzuki}}			
\affiliation{					
Department of Physics, Niigata University, Niigata 950-2181, Japan
} 						

\date{\today}

\begin{abstract} 				
Variational calculations for $s$-shell hypernuclei 
are performed by  explicitly including $\Sigma$ degrees of freedom. 
Four sets of $YN$ interactions 
(SC97d(S), SC97e(S), SC97f(S) and  SC89(S)) are used. 
The bound-state solution of \HeVL\ is obtained and a large energy 
expectation value of the tensor \LNSN\ transition part is found. 
The internal energy of the \HeIV\ subsystem is strongly affected by 
the presence of a $\Lambda$ particle with the strong tensor \LNSN\ 
transition potential. 
\end{abstract} 				


\maketitle 					

Few-body calculations for $s$-shell hypernuclei with mass number 
$A=3-5$ are important 
not only to explore exotic nuclear structure, including the  
strangeness degrees of freedom, but also to clarify the 
characteristic features of the hyperon-nucleon ($YN$) interaction. 
Although several interaction models 
have been proposed\cite{NSC89,NSC97,FSS96}, the detailed  
properties (e.g. $^1S_0$ or $^3S_1-^3D_1$ phase shift,  
strength of \LNSN\ coupling term) of the $YN$ interaction are 
different among the models. 
The observed separation energies (\BL's) of light $\Lambda$ 
hypernuclei  
are expected to provide important information on the $YN$ interaction, 
because the relative strength of the spin-dependent term or of the 
\LNSN\ coupling term is affected from system to   
system. 

Recently, few-body studies for $A=3,4$ hypernuclei 
have been conducted using modern $YN$ 
interactions\cite{Miyagawa,Hiyama,Nogga}. 
According to these developments, the Nijmegen soft core (NSC) model 
97f (or 97e)  
seems to be compatible 
with the experimental \BL's, 
though the calculated \BL\ for \HIVLs\ or \HeIVLs\ 
is actually slightly smaller than the experimental value. 
These few-body calculations, however, have not yet reached a stage 
to calculate \BL(\HeVL). 

If one constructs a phenomenological central $\Lambda N$ potential, 
which is consistent with the experimental \BL(\HIIIL), \BL(\HIVL), 
\BL(\HeIVL), \BL(\HIVLs) and \BL(\HeIVLs) values as well as the 
$\Lambda p$ total cross section, 
that kind of 
potential would overestimate the \BL(\HeVL) value\cite{DHT,Nemura}. 
This is 
known as an anomalously small binding of \HeVL. 
Though a suppression of the tensor forces \cite{Bando,Shinmura} or of 
the $\Lambda N-\Sigma N$ coupling \cite{Bodmer,Gibson} was discussed 
to be a   
possible mechanism to resolve the anomaly, the problem still remains 
an enigma\cite{SURVEY} due to the difficulty of performing a complete  
five-body treatment. 
Only one attempt was made, using a variational Monte Carlo 
calculation\cite{Carlson} with the NSC89 $YN$ interaction. 
Though NSC89 well reproduces both the experimental 
\BL(\HIIIL)\cite{Miyagawa} and \BL(\HIVL)\cite{Carlson,Nogga}  values   
as well as the experimental $\Lambda p$ total cross section, 
a bound-state solution of \HeVL\ was not found. 
In view of the aim to pin down a reliable $YN$ 
interaction, a systematic study for {\it all} $s$-shell hypernuclei  
is desirable. 

The $NN$ tensor interaction due to a one-pion-exchange mechanism is the 
most important ingredient for the binding mechanisms of light nuclei. 
More than a third, or about one half, of the interaction energy comes from 
the tensor force for the \HeIV\cite{ATMS,SVM,Pieper}. 
Since the pion-(or kaon-) exchange also induces the \LNSN\ transition for 
the $YN$ sector, both the $NN$ and \LNSN\ tensor interactions may also 
play important roles for light hypernuclei. 
If this is the case, the structure of the core nucleus (e.g. \HeIV) 
in the hypernucleus (\HeVL) 
would be strongly influenced by the presence of a $\Lambda$ particle.

The purpose of this letter is twofold: 
First is to perform an {\it ab initio} calculation for \HeVL\ 
as well as $A=3,4$ hypernuclei 
explicitly including $\Sigma$  degrees of freedom. 
Second is to discuss the structural aspects of \HeVL\ with an 
appropriate $YN$ interaction  
which is consistent with all of the $s$-shell hypernuclear data.

The Hamiltonian ($H$) of a system comprising nucleons and a hyperon  
($\Lambda$ or $\Sigma$) is given by $2\times 2$ components as 
\begin{equation}
H=\left(\begin{array}{cc}
H_\Lambda & V_{\Sigma-\Lambda} \\
V_{\Lambda-\Sigma} & H_\Sigma \end{array}
\right),\label{HTOT}
\end{equation}
where $H_\Lambda(H_\Sigma)$ operates on 
the $\Lambda$-($\Sigma$-)component and 
\begin{equation}
V_{\Lambda-\Sigma} = \sum_{i=1}^{A-1}v_{iY}^{(N\Lambda -N\Sigma)}. 
\label{VCC} 
\end{equation}
We employ the G3RS potential\cite{Tamagaki} for the $NN$ interaction 
and the SC97d(S), SC97e(S), SC97f(S) or SC89(S) potential\cite{Akaishi} 
for the $YN$ interaction, where all interactions have tensor and 
spin-orbit components in addition to the central one.   
We omit small { nonstatic correction} 
terms ($({\bfi L}\cdot{\bfi S})^2$ and ${\bfi L}^2$ terms) 
in the G3RS $NN$ interaction and odd partial-wave components in each 
interaction in order to focus on the main part of the interaction 
in the even parity state. 
The calculated binding energies for light nuclei 
(\HII, \HIII, \HeIII\ and \HeIV) are 2.28, 7.63, 6.98 and 24.57 MeV, 
respectively.  
The $YN$ interactions have Gaussian form 
factors whose parameters are set to reproduce
the low-energy $S$ matrix of the corresponding original Nijmegen $YN$ 
interactions\cite{ShinPrivate}.  
These Gaussian form factors help to save significant computer time.

The binding energies of various systems are calculated 
in a complete $A$-body treatment. 
The variational trial function must be flexible enough to  
incorporate both the explicit $\Sigma$ degrees of freedom and 
higher orbital angular momenta. 
The trial function is given by a combination of basis functions: 
\begin{equation}
\begin{array}{l}
\Psi_{JMTM_T} = \sum_{k=1}^{N} c_k \varphi_k, 
\quad \mbox{with}
\label{WFOFGS}\\
\varphi_k\!\!=\!\!{\cal A}\!\left\{G({\bfi x};A_k)
[\theta_{L_k}({\bfi x};u_k,K_k) \chi_{S_k}]_{JM} 
\eta_{kTM_T}\!\right\}. 
\label{DEFOFWF}
\end{array}
\end{equation}
Here, ${\cal A}$ is an antisymmetrizer acting on nucleons and 
$\chi_{S_k}$  $\left(\eta_{kTM_T}\right)$ is the spin (isospin) 
function. 
$\eta_{kTM_T}$ has two components: upper (lower) 
component refers to the $\Lambda$-($\Sigma$-)component. 
The abbreviation 
${\bfi x}\!=\!({\bfi x}_1, \cdots, {\bfi x}_{A-1})$ is
a set of relative coordinates. 
A set of linear variational parameters $(c_1,\cdots,c_{N})$ 
is determined by the Ritz variational principle.

A spatial part of the basis function is constructed by the
correlated Gaussian(CG) multiplied by the orbital angular momentum 
part $\theta_{L}({\bfi x})$, expressed by the global vector 
representation(GVR)\cite{GlobalV}.  
CG is defined by 
\begin{equation}
\begin{array}{rl}
G({\bfi x};A_k) & = \exp\Big\{-\frac{1}{2} 
    \sum_{i<j}^A{\alpha_k}_{ij} ({\bfi r}_i-{\bfi r}_j)^2\Big\} \\ 
& = \exp\Big\{-\frac{1}{2}
    \sum_{i,j=1}^{A-1}{(A_k)}_{ij}\,{\bfi x}_i\cdot{\bfi x}_j\Big\}. 
\end{array}
\label{DEFOFCORRG}
\end{equation}
The $(A-1)\times (A-1)$ symmetric matrix ($A_k$) is uniquely determined 
in terms of the interparticle correlation parameter ($\alpha_{kij}$). 
The GVR of $\theta_{L_k}({\bfi x};u_k,K_k)$ takes the form 
\begin{equation}
\begin{array}{l}
\theta_{L_k}({\bfi x};u_k,K_k)
= v_k^{2K_k+L_k}Y_{L_k}(\hat{\bfi v}_k), \quad \mbox{with}\\ 
{\bfi v}_k = \sum_{i=1}^{A-1} {(u_k)}_i{\bfi x}_i. 
\end{array}
\end{equation}
The $A_k$ and $u_k$ are 
sets of nonlinear parameters which 
characterize the spatial part of the basis function. 
Allowing the factor $v_k^{2K_k}$ $(K_k\ne 0)$ is useful 
to improve the short-range behavior of the trial function. 
The value of $K_k$ is assumed to take 0 or 1. 
The variational parameters are optimized by 
a stochastic procedure. 
The above form 
of the trial function gives accurate 
solutions. 
The reader is referred to Ref.\cite{SVM,GlobalV} for details 
and recent applications. 
For the spin and isospin parts, all possible configurations are 
taken into account.

\begin{table*}[] 
 \caption{$\Lambda$ separation energies, given in units of MeV, 
    of $A=3-5$ $\Lambda$-hypernuclei for different $YN$ interactions. 
    The scattering lengths, given in units of fm, 
    of  $^1S_0(a_s)$ and $^3S_1(a_t)$ states are also listed. 
  \label{BEOFSL}}
 \begin{ruledtabular} 					
  \begin{tabular}{lcccccccc}
      $YN$  & $a_s$ & $a_t$ &  	\BL(\HIIIL) & \BL(\HIVL) & \BL(\HIVLs) & 
                              \BL(\HeIVL) & \BL(\HeIVLs) & \BL(\HeVL)   \\
   \hline
      SC97d(S) & $-1.92$ & $-1.96$ & 
  	0.01 & 1.67 & 1.20 & 1.62 & 1.17 & 3.17 \\
      SC97e(S) & $-2.37$ & $-1.83$ & 
 	0.10 & 2.06 & 0.92 & 2.02 & 0.90 & 2.75 \\
      SC97f(S) & $-2.82$ & $-1.72$ & 
  	0.18 & 2.16 & 0.63 & 2.11 & 0.62 & 2.10 \\
      SC89(S) &  $-3.39$ & $-1.38$ & 
  	0.37 & 2.55 & unbound & 2.47 & unbound & 0.35 \\
   \hline
	Expt.  &  &  & 	$0.13\pm0.05$ & 
                        $2.04\pm0.04$ & $1.00\pm0.04$ & 
                        $2.39\pm0.03$ & $1.24\pm0.04$ & 
                        $3.12\pm0.02$ \\
  \end{tabular}
  \end{ruledtabular}
\end{table*}
Table~\ref{BEOFSL} lists the results of the $\Lambda$ separation 
energies. 
The scattering lengths of the $^1S_0(a_s)$ and $^3S_1(a_t)$ states for each  
$YN$ interaction are also listed in Table~\ref{BEOFSL}, where the 
interactions are given in increasing order of $|a_s|$ 
(and in decreasing order of $|a_t|$). 
The SC89(S) interaction produces no or very weakly bound state for 
\HIVLs, \HeIVLs\ or \HeVL.  
For the SC97d-f(S), the \BL(\HeVL) value is about $2-3$ MeV. 
This is a {\it first ab initio} calculation to produce the bound state 
of \HeVL\ with explicit $\Sigma$ degrees of freedom.  

The order of the spin doublet structure of the $A=4$ system
is correctly reproduced for all $YN$ interactions; 
the ground (excited) state has spin-parity, $J^\pi=0^+(1^+)$ 
for both isodoublet hypernuclei \HIVL\ and \HeIVL. 
Although the strengths of the $^1S_0$ and $^3S_1$ interactions of the 
SC97d(S) are almost the same as each other, the energy-level of the 
$0^+$ state is clearly lower than that of the $1^+$ state. 
All of the $A=3$ bound states given in Table~\ref{BEOFSL} have 
$J^\pi={1\over 2}^+$, in agreement with experiment. 
No other bound state 
has been obtained for all of the $YN$ interactions. 
For the SC97e(S), the differences between the calculated and 
experimental \BL\ values are the smallest among the $YN$ interactions 
employed in the present study.

\begin{table*}[] \centering \leavevmode 
 \caption{Probabilities, given in percentage, of finding a $\Sigma$ 
 particle in $A=3-5$ $\Lambda$-hypernuclei for different $YN$ 
 interactions.  
  \label{SIGPROB}}
 \begin{ruledtabular} 					
  \begin{tabular}{lcccccc}
	$YN$  & 
	$P_\Sigma$(\HIIIL) & $P_\Sigma$(\HIVL) & $P_\Sigma$(\HIVLs) & 
                             $P_\Sigma$(\HeIVL) & $P_\Sigma$(\HeIVLs) & 
        $P_\Sigma$(\HeVL) \\
   \hline
      SC97d(S) & 	0.06 & 1.27 & 1.37 & 1.24 & 1.35 & 2.04 \\
      SC97e(S) & 	0.15 & 1.49 & 0.98 & 1.45 & 0.96 & 1.55 \\
      SC97f(S) & 	0.23 & 1.88 & 1.09 & 1.83 & 1.08 & 1.87 \\
      SC89(S) & 	0.65 & 3.73 & unbound & 3.59 & unbound & 1.33 \\
  \end{tabular}
  \end{ruledtabular}
\end{table*}
Table~\ref{SIGPROB} lists the probability, \PS\ (in percentage), of 
finding a $\Sigma$ particle in the system. 
The sizable amount of \PS(\HeVL)'s is obtained. 
This implies that the $\Lambda-\Sigma$ coupling plays an important role, 
even for the \HeVL, despite a large excitation energy of the core nucleus, 
\HeIV\ (with the isospin 1), in the $\Sigma$-component. 
For the $A=4$ system, the \PS's of the $0^+$ state are about $1-2\%$, 
except for the SC89(S), 
while the \PS's of the $1^+$ state are nearly equal to or smaller than 
that of the $0^+$ state. 

\begin{figure}[]
 \includegraphics[width=.45\textwidth]{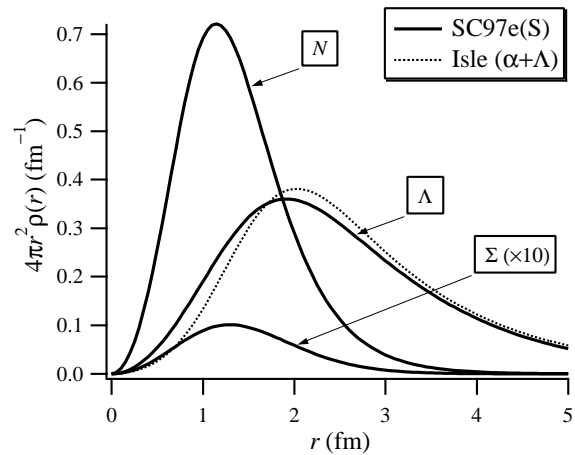}
 \caption{Density distributions of $N,\Lambda$ and $\Sigma$ for 
 \HeVL\ as a function of $r$, the distance from the center-of-mass of 
 \HeIV. 
 The SC97e(S) $YN$ interaction is used. 
 Note that the $\Sigma$-distribution has been multiplied by a factor of 
 $10$ to clarify the behavior. 
 The dotted line is taken from Ref.~\cite{Fuse}. 
 \label{DENSHeVL}}
\end{figure}

Figure~\ref{DENSHeVL} displays the density distributions for \HeVL\ 
using SC97e(S), and of $N,\Lambda$ and $\Sigma$ from the 
center-of-mass (CM) of \HeIV. 
Figure~\ref{DENSHeVL} also shows the $\Lambda$-distribution obtained 
from the Isle 
$\Lambda-\alpha$ potential\cite{Fuse}. 
The experimental pionic decay width of \HeVL\ suggests that 
the $\Lambda$-distribution should spread over a rather outer region   
compared to the distribution of the $\alpha$, 
as was discussed in Ref.~\cite{Fuse}. 
The present curve of the $\Lambda$-distribution is similar to that 
obtained by the Isle potential. 
The $\Sigma$-distribution has a shape similar to the $N$-distribution. 
The root-mean-square (rms) radii of $N$, $\Lambda$ and 
$\Sigma$ from the CM of the \HeIV\ are $1.5$, $2.9$ and $1.6$ fm,   
respectively. 

\begin{table*} \centering \leavevmode 
 \caption{Energy expectation values of the kinetic 
 and potential energy terms 
 for \HeVL, given in units of MeV. 
 The SC97e(S) $YN$ interaction is used. 
 For each potential part, 
 a summation over appropriate particle pairs is taken into account 
 (see Eq.~(\ref{VCC}) for  example) and 
 two central ($^1E$ and $^3E$) and a tensor 
 ($^3E$) components are listed separately.   
 The first (second) term of each element $\langle T_c\rangle$, 
 $\langle T_{Y-c}\rangle$ or $\langle V_{NN}\rangle$ represents 
 the first (second) term of Eq.~(\ref{OFDIAG}) 
 (${\cal O}=T_c,T_{Y-c}$ or $V_{NN}$). 
 The energy expectation values of $\langle T_c\rangle$ and three 
 $\langle V_{NN}\rangle$'s 
 for isolated \HeIV\ are $84.86, -33.22, -33.05$ and $-43.93$ MeV, 
 respectively.  
 \label{INSHeVL}}
 \begin{ruledtabular} 					
  \begin{tabular}{ccccrrrr}
    & $\langle T_c\rangle$ & $\langle T_{Y-c}\rangle$ & 
   $\langle V_{NN}\rangle$ & 
   $\langle V_{N\Lambda}\rangle$ & 
   $2\langle V_{\Lambda-\Sigma}\rangle$ & 
   $\langle V_{N\Sigma}\rangle$ & 
   \\
   \hline
            & ${83.43}{+2.74}$ & $9.11+3.88$ & $-33.14-0.35$ & $-3.97$ & $- 0.02$ & $0.07$ & (Central, $^1E$) \\
            &              &             & $-32.03-0.27$ & $ 2.98$ & $- 1.02$ & $1.56$ & (Central, $^3E$) \\
            &              &             & $-40.91-0.12$ & $-2.24$ & $-19.51$ & $0.87$ & (Tensor, $^3E$) \\
\end{tabular}
  \end{ruledtabular}
\end{table*}
Table~\ref{INSHeVL} lists the energy expectation values of the kinetic 
and potential energy terms for \HeVL. 
The contributions from the spin-orbit and the Coulomb potentials 
are not shown in the table, though the calculations include them. 
Here, $T_c$ is the kinetic energy of the core nucleus ($c$) 
subtracted by the CM energy of $c$: 
\begin{equation}
T_c = \sum_{i=1}^{A-1}{{\bfi p}_i^2\over 2m_N} 
 - {\left(
\sum_{i=1}^{A-1}{\bfi p}_i
\right)^2
\over 2(A-1)m_N}. 
\label{TC}
\end{equation}
The kinetic energy of the relative motion between the $Y$ and the CM of 
$c$ is given by 
\begin{equation}
T_{Y-c} = {{\bfi \pi}_{Y-c}^2\over 2\mu_Y}+\! \left(m_Y\!-\!m_\Lambda \right)\!c^2, 
\label{TYC}
\end{equation}
where $\mu_Y\!=\!{(A-1)m_N m_Y\over (A-1)m_N+m_Y}$ is the reduced mass for 
the $Y\!+c$ system and ${\bfi \pi}_{Y-c}$ is the canonical momentum of 
the relative coordinate between $Y$ and $c$ $(Y=\Lambda,\Sigma)$.  
$T_{Y-c}$ also counts the difference in the rest-mass energy between
$\Lambda$ and $\Sigma$.  
Each potential part $\langle V\rangle$ takes account of a summation over 
appropriate particle pairs (see Eq.~(\ref{VCC}) for example). 
The energy expectation values of the first three columns in 
Table~\ref{INSHeVL} are written as 
\begin{equation}
 \begin{array}{rcl}
  \langle{\cal O}\rangle &=& \langle\Psi_\Lambda|{\cal O}|\Psi_\Lambda\rangle + \langle\Psi_\Sigma|{\cal O}|\Psi_\Sigma\rangle, 
 \end{array}
 \label{OFDIAG}
\end{equation}
where the upper (lower) component of the $\Psi_{JMTM_T}$ is denoted by 
$\Psi_\Lambda$ ($\Psi_\Sigma$). 
The first (second) term of each element 
($\langle T_c\rangle$, $\langle T_{Y-c}\rangle$ or $\langle V_{NN}\rangle$) 
in Table~\ref{INSHeVL} represents the first (second) term of  
Eq.~(\ref{OFDIAG}). 
The 
energy of the \HeIV\ subsystem changes a lot 
from that of the isolated one, 
\begin{eqnarray}
\Delta E_c &=& \Big(\langle T_c\rangle + \langle V_{NN}\rangle \Big)_{\mbox{\HeVL}}  
- \Big(\langle T_c\rangle + \langle V_{NN}\rangle \Big)_{\mbox{\HeIV}} \nonumber \\ 
&\approx& 4.7 \mbox{MeV}. 
\end{eqnarray}
This difference is considerably large despite the fact that the rms 
radius of $N$  
from the CM of \HeIV\ for the \HeVL\ hardly changes from that for  
\HeIV. (Both radii are 1.5~fm.) 
Most of the change is due to a reduction of the energy expectation 
value of the tensor $NN$ interaction, 
$$
\Big(\langle V_{NN}(\mbox{tensor})\rangle \Big)_{\mbox{\HeVL}}  
- \Big(\langle V_{NN}(\mbox{tensor})\rangle \Big)_{\mbox{\HeIV}} 
\approx 2.9 \mbox{MeV}. 
$$
On the other hand, the tensor \LNSN\ transition part has 
a surprisingly large energy expectation value (about $-20$ MeV). 
This large coupling energy makes \HeVL\ bound in spite of both the 
energy loss of $\Delta E_c$ and the extremely 
high energy of the $\Sigma$-component 
(${\langle H_\Sigma\rangle\over P_\Sigma}\sim 600$ MeV).  

The calculated wave function is divided 
into 
orthogonal components 
according to the total orbital angular momentum ($L$), 
the total spin ($S$), the core nucleus spin ($S_c$) and 
the core nucleus isospin ($T_c$).  
Table~\ref{PROBHeVL} displays the probability 
of each component for \HeVL. 
The table also lists the probability of $S$-state or of $D$-state for 
\HeIV. 
The sizable amount of probability of the $\Sigma$-component is found 
in the $D$-state while the sum of $D$-state probabilities in the 
$\Lambda$-component  
is slightly smaller than that for \HeIV. 
Moreover, though the presence of a $\Lambda$ in \HeIV\ with the 
strong tensor \LNSN\ transition potential influences the structure of 
the $D$-state component and reduces the energy expectation 
value of the tensor $NN$ interaction, the large coupling energy 
$\langle V_{\Lambda-\Sigma}\rangle$ of the tensor part bears the bound 
state of \HeVL\ instead. 
\begin{table}[] \centering \leavevmode 
 \caption{Probability, given in percentage, 
 of each component with the 
 total orbital angular momentum ($L$), total spin ($S$), core nucleus 
 spin ($S_c$) and core nucleus isospin ($T_c$) in $\Lambda$- or in 
 $\Sigma$-component for \HeVL.  
 The SC97e(S) $YN$ interaction is used. 
 The probability in $S$- or in $D$-state for \HeIV\ is also listed.
  \label{PROBHeVL}}
 \begin{ruledtabular} 					
  \begin{tabular}{lccccccc}
   & \multicolumn{2}{c}{$L=0$} & & \multicolumn{4}{c}{$L=2$} \\ 
   \cline{5-8}
   & \multicolumn{2}{c}{$S={1\over 2}$} & &
   \multicolumn{2}{c}{$S={3\over 2}$} & & $S={5\over 2}$ \\
   \cline{2-3} \cline{5-6} \cline{8-8}
   & $S_{c}=0$ & $S_{c}=1$ & & $S_{c}=1$ & $S_{c}=2$ & & 
   $S_{c}=2$ \\ 
   \hline
   \HeVL \\
   $(T_c=0)\otimes\Lambda$ & 89.14 & 0.03 & & 0.19 & 3.74 & & 5.36 \\
   $(T_c=1)\otimes\Sigma$  & 0.10 & 0.09 & & 1.34 & $\sim 0$ & & 0.01 \\
   \hline
   $^4$He & 89.56 & & & & \multicolumn{3}{c}{10.44} \\
  \end{tabular}
  \end{ruledtabular}
\end{table}

In summary, we have made a systematic study of all $s$-shell 
hypernuclei based on {ab initio} 
calculations 
using $YN$ interactions with an explicit $\Sigma$ admixture. 
The bound-state solution of \HeVL\ was obtained. 
As the Ref.~\cite{Nogga} claimed, though there is none of the 
interaction 
models to 
describe very precisely the experimental \BL's, the five-body 
calculation convinced us that the anomalous binding problem would be 
resolved by taking account of the explicit $\Sigma$ admixture. 
The \BL\ values, obtained by using SC97e(S), are the closest to the 
experimental values, among the $YN$ interactions employed in this study.  
A sizable amount of \PS\ was obtained, even for the \HeVL, in spite of 
the large excitation energy of \HeIV. 
The contribution of the energy from the tensor \LNSN\ coupling is quite 
large, and this  
coupling is considerably important to make \HeVL\ bound. 
This is novel finding in contrast with the Brueckner-Hartree-Fock 
calculation~\cite{Akaishi}. 
The present study for \HeVL\ is a first step toward 
a detailed description of light strange nuclear systems. 
The core nucleus, \HeIV, is no longer rigid 
in interacting with 
a $\Lambda$ particle. 
A similar situation can occur for strangeness $S=-2$ systems. 
Investigations into the strength of the $\Lambda\Lambda$ interaction 
based on the experimental data of the binding energy for double $\Lambda$ 
hypernuclei (e.g. \HeVILL~\cite{Nagara}) should take 
account of the energy reduction of the core nucleus ($\Delta E_c$ for 
\HeVILL\ is expected to be larger than the present 
$\Delta E_c$ for \HeVL). 

\begin{acknowledgments}         
We are thankful to Y. Fujiwara, K. Miyagawa, H. Kamada, 
K. Varga and S. Shinmura for useful communications. 
One of the authors (H.N.) would like to thank for 
JSPS Research Fellowships for Young Scientists. 
The calculations of $A=4,5$ systems were made using the RCNP's SX-5 
 computer and KEK's SR8000 computer. 
\end{acknowledgments}          

\end{document}